# Extending the Defect Tolerance of Halide Perovskite Nanocrystals to Hot Carrier Cooling Dynamics

Junzhi Ye [1,2#], Navendu Mondal [3#*], Ben P. Carwithen [3], Yunwei Zhang [4], Linjie Dai [1,5], Xiangbin Fan [6], Jian Mao [5,7], Zhiqiang Cui [4], Pratyush Ghosh [1], Clara Otero-Martínez [8], Lars van Turnhout [1], Zhongzheng Yu [1], Ziming Chen [3], Neil C. Greenham [1], Samuel D. Stranks [1,5], Lakshminarayana Polavarapu [8], Artem Bakulin [3], Akshay Rao [1], Robert L.Z. Hoye [2,9*]

1. Cavendish Laboratory, University of Cambridge, 11880, Cambridge CB3 0HE, United Kingdom
2. Inorganic Chemistry Laboratory, University of Oxford, South Parks Road, Oxford OX1 3QR, United Kingdom
3. Department of Chemistry and Centre for Processable Electronics, Imperial College London, Molecular Sciences Research Hub, 83 Wood Lane, London W12 0BZ, United Kingdom
4. School of Physics, Sun Yat-sen University, 510275 Guangzhou, China
5. Department of Chemical Engineering and Biotechnology, University of Cambridge, Cambridge CB3 0AS, United Kingdom.
6. Department of Engineering, University of Cambridge, 9 JJ Thomson Avenue, Cambridge, CB3 0FA, United Kingdom
7. State Key Laboratory of Photovoltaic Science and Technology, Shanghai Frontiers Science Research Base of Intelligent Optoelectronics and Perception, Institute of Optoelectronics, Fudan University, Shanghai, 200433, China
8. CINBIO, Universidade de Vigo, Materials Chemistry and Physics Group, Department of Physical Chemistry, Campus Universitario As Lagoas, Marcosende, 36310 Vigo, Spain
9. Department of Materials, Imperial College London, Exhibition Road, London SW7 2AZ, United Kingdom.

\# These authors contributed equally to this work.

\* Email: n.mondal@imperial.ac.uk (N.M.), robert.hoye@chem.ox.ac.uk (R. L. Z. H.)

**Abstract**

Defect tolerance is a critical enabling factor for efficient lead-halide perovskite materials, but the current understanding is primarily on band-edge (cold) carriers, with significant debate over whether hot carriers (HCs) can also exhibit defect tolerance. Here, this important gap in the field is addressed by investigating how internationally-introduced traps affect HC relaxation in $CsPbX_3$ nanocrystals (X = Br, I, or mixture). Using femtosecond interband and intraband spectroscopy, along with energy-dependent photoluminescence measurements and kinetic modelling, it is found that HCs are not universally defect tolerant in $CsPbX_3$, but are strongly correlated to the defect tolerance of cold carriers, requiring shallow traps to be present (as in $CsPbI_3$). It is found that HCs are directly captured by traps, instead of going through an intermediate cold carrier, and deeper traps cause faster HC cooling, reducing the effects of the hot phonon bottleneck and Auger reheating. This work provides important insights into how defects influence HCs, which will be important for designing materials for hot carrier solar cells, multiexciton generation, and optical gain media.

**Introduction**

The power conversion efficiency (PCE) of lead iodide-based perovskite photovoltaics (PVs) has now reached a certified value of 26.2% under 1-sun illumination[1], which is rapidly approaching the radiative PCE limit of ~30%[2,3]. This radiative limit in efficiency comes from the inefficient management of heat dissipation, especially for charge-carriers excited above the bandgap (*i.e.*, hot carriers, or HCs)[4-9]. However, practically, the excess energy of the HCs cannot be fully utilized and extracted efficiently because HC cooling is much faster than charge-carrier extraction[9-11]. Therefore, understanding the origin of the fast-cooling process in halide perovskites with different bandgaps and compositions is crucial for developing approaches to sufficiently slow down HC cooling, such that these HCs can be collected. In general, HCs release this excess energy through ultrafast (<100 fs) carrier-carrier scattering, followed by carrier-phonon interaction events (<1 ps) to reach the lattice temperature. Moreover, studying HC dynamics provides a fundamental understanding of the carrier-carrier and carrier-phonon interactions in these materials with implications for charge-carrier transport and material stability[12,13].

One of the key enabling, but unusual, properties of lead halide perovskites (LHPs) is their reported defect tolerance, meaning that the lifetimes and mobilities of free charge-carriers are relatively insensitive to the presence of defects[14-17]. However, defect tolerance has generally been discussed with regard to band-edge (*i.e.*, "cold") carriers undergoing non-radiative recombination, and is typically quantified by PLQY and time-resolved PL measurements. A critical question is whether this defect tolerance extends to HCs. Bonn *et al.* suggested this to be so[18], whereas Jiang *et al.* suggested that while band-edge carriers in MAPbI$_3$ were defect-tolerant, the HC lifetime was shortened due to trapping at the grain boundaries of polycrystalline films[19]. HC-activated trapping[20,21], along with passivation-induced enhancement of HC lifetime[9,12,21-25] and transport[26], was also observed in a few other instances. Righetto *et al.* reported that the HC cooling process can be directly affected by defect trapping on the surface of perovskite nanocrystals (NCs)[21]. Zhou *et al.* utilized density functional theory (DFT) calculations to suggest that common point-defects, such as iodide and methylammonium vacancies in MAPbI$_3$ perovskite, can adversely affect the HC lifetime[9]. The defect tolerance of HCs is therefore not known, with contradictory evidence in the current literature. Our work aims to address this by systematically establishing the relationship between HC cooling lifetime and defect densities and energies in compositionally-tunable perovskites. This will be important for developing wide-gap and narrow-gap perovskite absorbers for single-junction and tandem solar cells, and could show whether HC perovskite solar cells are attainable[27,28].

To elucidate the role of traps on HC cooling lifetime, $CsPbX_3$ NCs were selected as the material system of interest rather than LHP thin films, due to the capability of introducing different defect densities intentionally by controlling the NC surface chemistry through multiple antisolvent purification steps[29]. This enables a more direct correlation between the effects of defects and non-radiative recombination, which affects the photoluminescence quantum yield (PLQY). By progressively increasing the defect density, we employed femtosecond pump-probe (PP) and pump-push-probe (PPP) transient absorption (TA) spectroscopy to investigate the effect of trap density and energy on the HC cooling dynamics. The TA spectroscopy and kinetic modelling on HC cooling dynamics demonstrate that HC lifetime is governed by both defect density and energy, which are in turn controlled by composition and bandgap. We find that HCs are protected in narrow-bandgap perovskite NCs comprising shallow traps compared to those in wide-gap NCs. By demonstrating the defect-tolerant nature of HCs in perovskites, design guidelines are provided that can pave the way for efforts to develop HC solar cells.

**Intentionally tuning trap densities, and the effects on cold carriers**

We synthesised colloidally stable $CsPbX_3$ (X = Br, I) NCs by hot-injection (see Methods for details)[29,30]. To intentionally introduce different defect densities, the as-synthesised $CsPbBr_3$, $CsPbBr_xI_{3-x}$ and $CsPbI_3$ NCs were purified multiple times using the low-polarity antisolvent methyl acetate. This process results in the partial removal of surface ligands and halides without affecting the size and structure of the NCs, as shown in Fig. 1a and Supplementary Fig. 1 [29,31]. From X-ray photoemission spectroscopy (XPS) measurements (Supplementary Fig. 2a-c), there was a decrease in the surface halide to Pb ratio with an increase in the number of purification steps, indicating an increase in the density of surface halide vacancies. Following purification, the absorption and PL spectra of the pure halide NCs did not exhibit any significant changes (Fig. 1b and d), while the mixed-halide perovskite NCs significantly blue-shifted (Fig. 1c). This was due to selective etching (expulsion) of surface iodine species, resulting in the bandgap widening[29]. The decrease in the PLQY (Fig. 1) and PL lifetime (Supplementary Fig. 3a,b) of the mixed-halide and bromide perovskites with an increase in the number of purification steps (Fig. 1a) is consistent with the introduction of traps due to halide vacancies[31-33]. By contrast, there was very little change in the PLQY and PL lifetime of the narrow-gap $CsPbI_3$ following purification (Fig. 1a, Supplementary Fig. 3c), and this is consistent with the greater defect tolerance of $CsPbI_3$ due to most iodide vacancies being shallow.

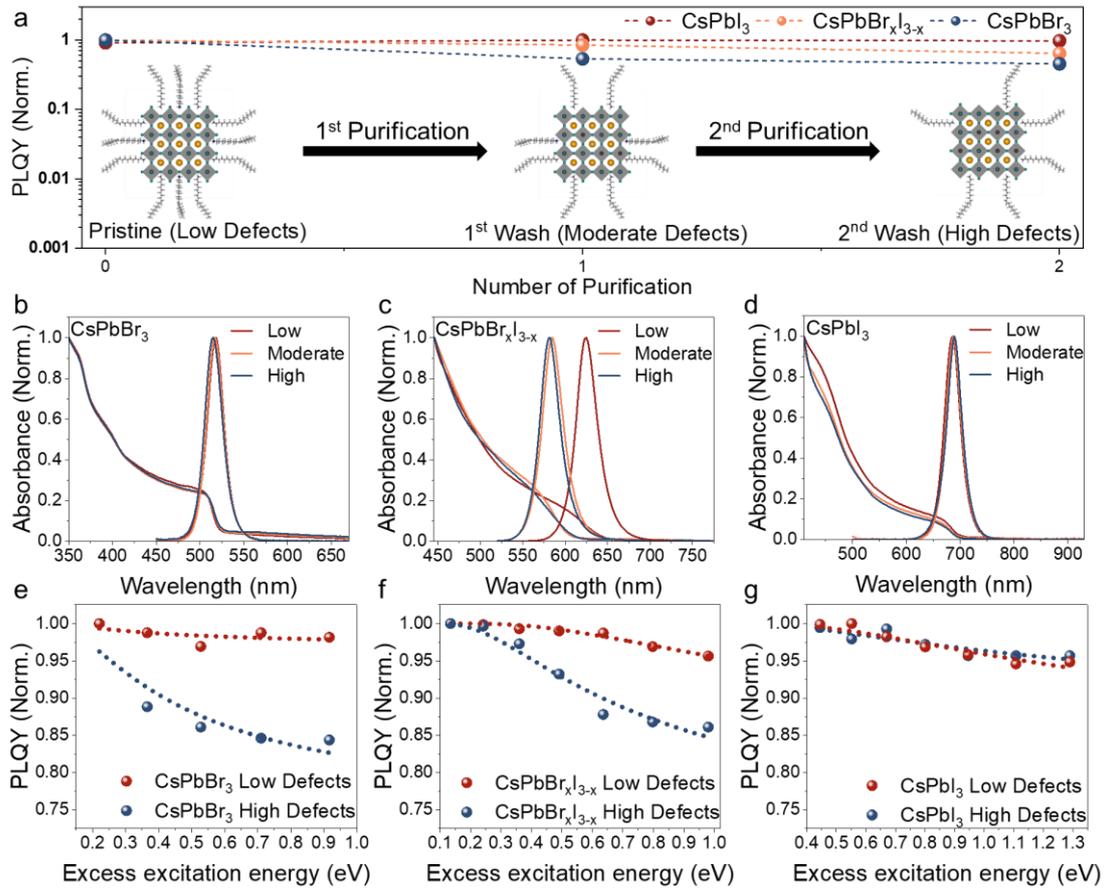

**Fig. 1 | Effect of intentionally introduced defect states on the optoelectronic properties of CsPbX$_3$ nanocrystals (X = I, Br). a,** Change in the normalized photoluminescence quantum yield (PLQY) of CsPbI$_3$, CsPbBr$_x$I$_{3-x}$ and CsPbBr$_3$ nanocrystals after successive purification steps. All PLQY values were normalised at the lowest excess energy. Comparison of the photoluminescence (PL) and absorption spectra of pristine (low defect density), singly purified (moderate defect density), and doubly purified (high defect density) **b,** CsPbBr$_3$, **c,** CsPbBr$_x$I$_{3-x}$ and **d,** CsPbI$_3$ nanocrystals. Excitation-wavelength-dependent PLQY for **e,** CsPbBr$_3$, **f,** CsPbBr$_x$I$_{3-x}$, and **g,** CsPbI$_3$ nanocrystals. In each case, a comparison is made between nanocrystals with low and high defect densities. The model fit to the excitation-dependent measurements was obtained from Ref. 21.

The PLQY, PL lifetime and XPS measurements suggest that we have successfully prepared NCs with different defect densities. To gather preliminary insight into the role of defects on the dynamics of HCs, we first performed excitation-energy-dependent PLQY measurements. These measurements (Fig. 1e-g) show that for less defective NCs, the PLQY remained relatively constant with changes in excitation energy. By contrast, for more defective (*i.e.*, doubly purified) CsPbBr$_3$ and CsPbBr$_x$I$_{3-x}$ NCs, the PLQY decreased by ~15% as the excitation energy increased by ~1 eV above the bandgap (Fig. 1e,f). If the PLQY were solely dependent on the recombination of band-edge carriers, we would expect the PLQY to remain the same irrespective of the excitation energy. Indeed, we observe only a small change in PLQY with increasing excess energy in CsPbI$_3$ NCs, and this was not exacerbated by introducing a higher defect density (Fig. 1g). Together, these results point to the conclusion that when the excitation energy is high (w.r.t. the bandgap) in wide-gap perovskites, HCs undergo additional non-radiative recombination pathways, thus lowering PLQY.

We now utilize a simplified HC trapping model originally developed by Righetto *et. al.*[21] (detailed in Supplementary Note 1 and Supplementary Fig. 4) to explain the HC trapping process. These fits are shown as the dotted line in Fig. 1e-g, and the details about the model are described in Supplementary Note 1 and Supplementary Fig. 4, with fitting parameters shown in Supplementary Table 1. The fitted values suggest that the high defect-density samples tend to have lower radiative recombination constants and greater electronic coupling to traps than the low defect-density samples, and this effect is more significant for wider-gap systems. The apparent 'tolerance' of HCs to traps in CsPbI$_3$ could be due to these traps being shallower than in CsPbBr$_3$ and CsPbBr$_x$I$_{3-x}$ NCs, such that the overlap between the conduction band and trap states is smaller with a smaller energy offset for the HCs to be trapped[21]. Consistent with this proposed explanation, we verified from DFT calculations that the halide vacancy in CsPbI$_3$ (0.278 eV from the conduction band minimum) is much shallower compared to Br/I-based (0.513 eV) and Br-based (0.666 eV) NCs (Supplementary Fig. 5). The absolute values of these trap states might not be completely accurate, but the trend in defect positions match with the literature [34,35].

Following this qualitative evidence, we now turn our attention to characterizing the effects of defects through time-resolved spectroscopic techniques, namely two-pulse transient absorption spectroscopy (TAS). Fig. 2a and b show the TA spectra (within 0.5-30 ps) of the low and high defect density CsPbBr$_3$ samples. There are three spectral signatures: i) a high-energy, broad negative photo-induced absorption (PIA), ii) positive ground state bleach (GSB), which we attribute to the depopulation of band-edge carriers, and iii) a negative sub-bandgap short-lived PIA (red-side of GSB), which is often ascribed to band-gap renormalisation (noting that our labelling for Fig. 2a,b is in $\Delta T/T$). This sub band-gap PIA would disappear upon band-edge excitation and the decay dynamics are correlated with the formation of GSB (due to state-filling processes). As we have intentionally introduced defects into the NCs, the decay dynamics of the sub-bandgap region are a combination of a positive signal due to trap bleach and a negative signal due to bandgap renormalisation. Fig. 2c shows the kinetics at 535-545 nm wavelength (sub-bandgap), which displays a clear difference between the low and high defect density NCs. The positive $\Delta T/T$ signal observed from high defect density NCs in the sub-bandgap wavelength range is assigned to the trap-induced bleach (TB) signal, as shown in Fig. 2c and f, indicating that there are charge-carriers being trapped into the defect states even during the HC cooling time window (0-10 ps), and the bleach signal from trapped carriers slowly decayed away after a few picoseconds to sub-nanoseconds, as shown by the blue data points for high defect density NCs in Fig. 2c. Similarly, a higher $\Delta T/T$ signal for sub-bandgap decay is observed from mixed halide NCs for the high defect density NCs (Fig. 2d), but no obvious change is observed for pure I-based NCs (Fig. 2e). To make a careful assessment of the sub-band-gap region TA signal comprising the spectrally overlapping feature of PIA and trap-bleach, we have performed global analysis (results shown in Supplementary Fig. 6), which clearly resolves their individual contributions and is consistent with our analysis presented in Figure 2. Interestingly, a positive component is identified after deconvolution for the defective $CsPbBr_xI_{3-x}$ NCs due to trap bleaching, but which is not apparent for in the $CsPbI_3$ NCs. This is consistent with our energy-dependent PLQY measurements, suggesting that HCs in I-based NCs are less influenced by defects.

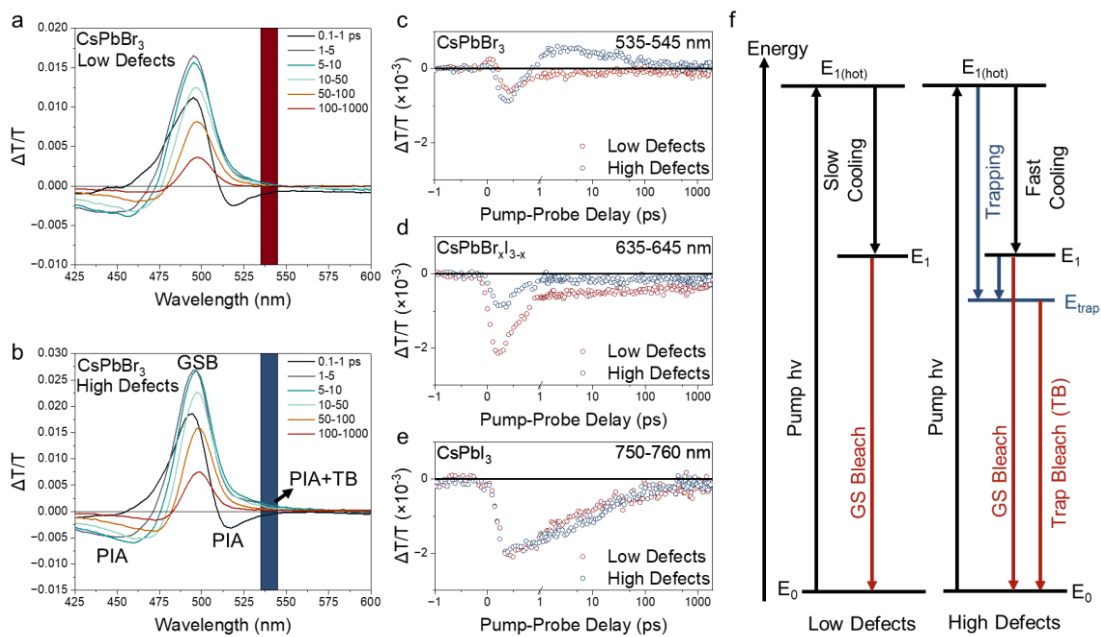

**Fig. 2 | Evidence of carrier trapping from pump-probe transient absorption spectroscopy. a, b,** Transient absorption spectra of the singly-purified (low defect density) and doubly-purified $CsPbBr_3$ NCs (high defect density). Red and blue regions indicate the ground state bleach (GSB) (490-500 nm) and sub-bandgap trap bleach (TB) (535-545 nm) regions that were integrated to determine the kinetics. The kinetics of the TB for singly (low defect density)- and double-purified (high defect density) **c,** $CsPbBr_3$ NCs (probed at 535-545 nm), **d,** $CsPbBr_xI_{3-x}$ NCs (probed at 635-645 nm), and **e,** $CsPbI_3$ NCs (probed at 750-760 nm). The TA measurements were performed at 116.6 μJ cm$^{-2}$ pulse$^{-1}$ fluence, and the spectra are shown in Supplementary Fig. 7-9. **f.** Schematic representation of the charge-carrier relaxation processes highlighting the carrier trapping events, with energy levels indicated here are arbitrarily. The NC solution was pumped with photon energies significantly higher than the bandgap ($\hbar\omega_{pump}$ = 3.1 eV).

**Influence of traps on hot carrier lifetime probed by two-pulse TAS**

Both excitation-energy-dependent PLQY and the sub-bandgap states of the TA analysis suggest that defects can influence the HC cooling process for wide-bandgap systems based on indirect monitoring of the band-edge carriers after cooling. In this section, we will discuss the effect of defects directly on the cooling dynamics and HC lifetime ($\tau_{cool}$), defined as the time required for HCs to equilibrate with the surrounding lattice temperature, using two-pulse TAS. Fig. 3a-f shows the 2D color map of TA spectra for all of the systems with increasing defect densities measured at high fluence. It is clearly evident that the GSB signals of the $CsPbBr_3$ and $CsPbBr_xI_{3-x}$ NCs become narrower with an increase in the defect densities in these systems. The narrowing of the GSB is an indication of fast HC cooling. However, the GSB of $CsPbI_3$ NCs did not show any noticeable narrowing as we increased the defect density (Fig. 3g-i).

To further investigate the trapping effect on the kinetics of HC cooling, we extracted cooling temperature and lifetimes via the widely used procedure of fitting a Maxwell–Boltzmann distribution (Eq. 1) to the high-energy GSB tail, by approximating it from the Fermi-Dirac distribution of HCs[10,36] (Eq. 1) as shown in Supplementary Figs. 7-9:

$$-\frac{\Delta T}{T}(E) \propto e^{-\frac{E-E_f}{k_B T_c}} \quad (1)$$

where $\frac{\Delta T}{T}(E)$ is the transient transmittance obtained from the TA measurement, $T_c$ is the carrier temperature. The energy loss rate ($J_{loss}$) in eV ps$^{-1}$ can be estimated for NCs with different compositions and defect densities based on Eq. 2[37]:

$$J_{loss} = -\frac{3}{2}\frac{k_B dT_c}{dt} \quad (2)$$

The results are shown in Fig. 3j-l. For Br-based NCs, $J_{loss}$ increased from around 0.03 eV ps$^{-1}$ to 0.2 eV ps$^{-1}$ upon defect introduction at a carrier temperature of 1000 K (Fig. 3j). Similarly, $J_{loss}$ increased from around 0.008 eV ps$^{-1}$ to 0.04 eV ps$^{-1}$ for low to moderate defect densities, and to 0.17 eV ps$^{-1}$ for high defect densities in Br/I-based NCs at 1000 K (Fig. 3k). In contrast, $J_{loss}$ remained relatively constant for I-based NCs, which is around 0.13 eV ps$^{-1}$ at 1000 K. This suggests that the presence of defects introduces an additional energy loss pathway in wider bandgap NCs, and thus the loss rate is higher for high defect density systems. However, these defects are less influential for the low-gap $CsPbI_3$, since the energy loss rate did not change drastically for these NCs.

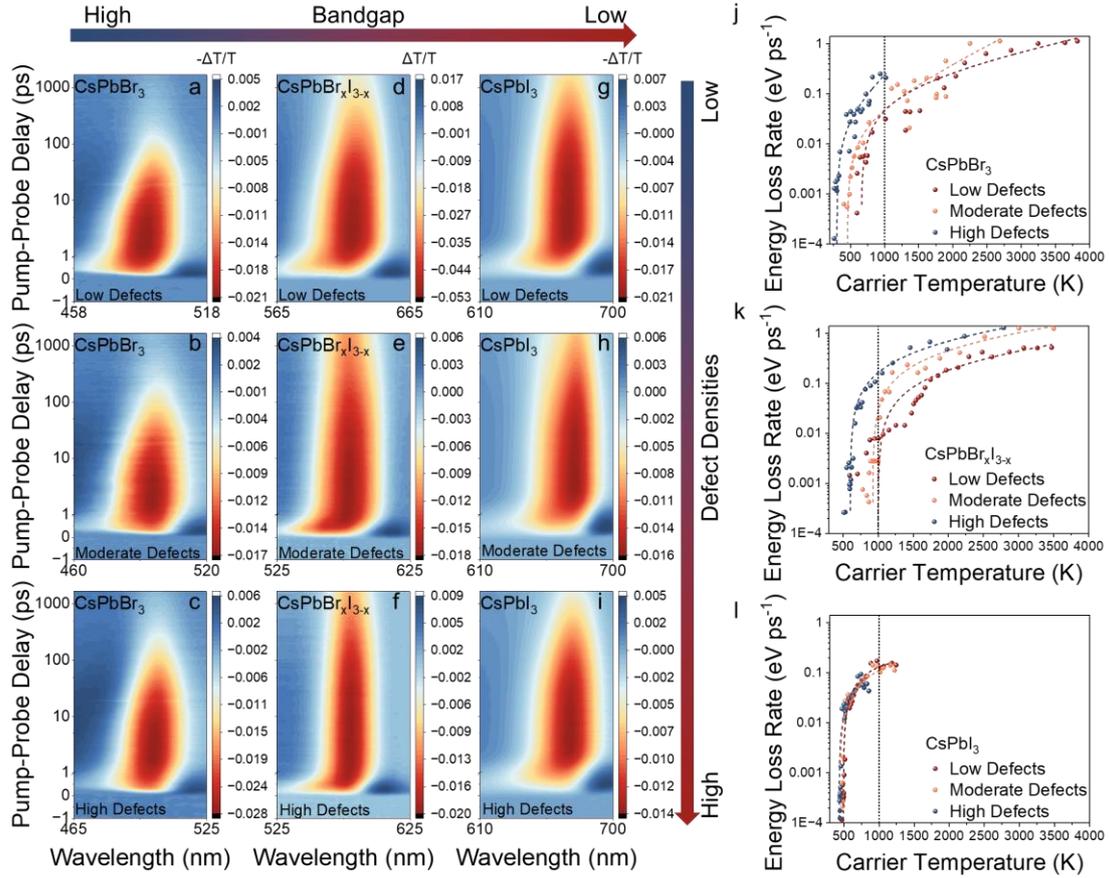

**Fig. 3 | Hot carrier energy loss rate based on pump-probe transient absorption spectroscopy.** TA maps for low defects, moderate defects, and high defects of **a-c,** $CsPbBr_3$, **d-f,** $CsPbBr_xI_{3-x}$ and **g-i,** $CsPbI_3$ perovskite nanocrystal solutions measured under 400 nm wavelength excitation. Energy loss rate for different defect concentrations in **j,** $CsPbBr_3$, **k,** $CsPbBr_xI_{3-x}$, and **l,** $CsPbI_3$ NCs. In all cases, a comparison between pristine (low defects, red), singly purified (moderate defects, yellow) and doubly purified NCs (high defects, blue) was made shown in the vertical dotted line. The energy loss rate is based on the relaxation lifetime at 194.3 μJ cm$^{-2}$ pulse$^{-1}$ for $CsPbBr_3$ (carrier density of 12.5×10$^{17}$ cm$^{-3}$), 178.2 μJ cm$^{-2}$ pulse$^{-1}$ for $CsPbBr_xI_{3-x}$ (carrier density of 43.5×10$^{17}$ cm$^{-3}$) and 193.4 μJ cm$^{-2}$ pulse$^{-1}$ for $CsPbI_3$ (carrier density of 16.6×10$^{17}$ cm$^{-3}$).

The extracted charge-carrier temperature as a function of time delay for different initial carrier densities or fluences for NCs with different compositions are presented in Supplementary Fig. 7 to 9h-g. For low fluence, the time-dependent changes in carrier temperatures can be fit with a single exponential decay, as shown in in Supplementary Tables 2-4, corresponding to rapid carrier cooling via Fröhlich interactions, in which scattering between charge-carriers and longitudinal optical (LO) phonons is the dominant pathway at room temperature[10]. At moderate excitation density, the HC lifetime gradually increases as carriers compete for a finite availability of phonon modes into which they may transfer their excess energy – a phenomenon known as the hot phonon bottleneck[38,39]. Apart from phonon heating, enhanced carrier-carrier interactions at high carrier concentrations could also contribute to prolonged HC lifetimes through the non-radiative Auger heating process. Here, the energy released upon recombination of an electron-hole pair is transferred to a neighbouring carrier, thus replenishing the population of HCs. Unlike bulk systems, this Coulombic interaction-mediated Auger heating process often dominates in quantum confined NCs. Thus, at high carrier density, the appearance of an additional slower component after the initial sub-ps decay is assigned to Auger heating in our NC-based systems. As a result, biexponential fitting is required to capture the whole cooling trajectory, in which the fast and slow lifetime components (under moderate to high carrier densities) can be ascribed to the hot phonon bottleneck and Auger heating effect, respectively, and the fitted results are listed in Supplementary Table 2-4. Interestingly, this second slow cooling component at high fluence became less obvious for the moderately-defective and highly-defective NCs. To give an example, for Br-based systems, at a carrier density of $12.5 \times 10^{17}$ cm$^{-3}$, the HC cooling lifetime decreases from $3.84 \pm 0.34$ ps (for the low defect density sample), to $2.63 \pm 0.41$ ps and $0.96 \pm 0.07$ ps for the moderate and high defect density samples, respectively (Fig. 4a and Supplementary Table 2), due to the substantial decrease in the $t_2$ component, which is strongly influenced by excitation fluence. This indicates that the presence of higher defect densities introduces additional relaxation routes (*i.e*., trap-mediated) to the intrinsic intraband relaxation process, therefore leading to a reduction in the cooling time over the entire carrier density range studied here. The same results are shown in Supplementary Fig. 8h-g for CsPbBr$_x$I$_{3-x}$ NCs, where the second component appeared clearly at a carrier density of $58.0 \times 10^{17}$ cm$^{-3}$ or higher, and almost disappeared for the moderate and high defect density samples. The cooling lifetime decreased from $7.05 \pm 0.05$ ps (for the low defect density sample) to $4.76 \pm 0.18$ ps to $4.46 \pm 0.04$ ps for the moderate to high defect density sample (Fig. 4b and Supplementary Table 3). However, for the CsPbI$_3$ NCs, the change of the spectra, map and extracted carrier cooling kinetics did not vary significantly with the number of purification steps (Supplementary Fig. 9). The HC lifetime was found to be similar between the pristine and defective samples (around 12-13 ps at the carrier density of $33.3 \times 10^{17}$ cm$^{-3}$, as shown in Fig. 4c and Supplementary Table 4). In summary, for wider bandgap systems (Br and Br/I-based NCs), $\tau_{cool}$ is shorter as the defect density increased, but this phenomenon is less obvious for lower bandgap I-based NCs (Fig. 4d). We also observed that the differences in the extracted HC lifetimes become smaller as the carrier density increases due to the trap filling effect. When the traps are occupied with the large density of carrier (generated by pump laser of higher fluence), the additional loss pathway for the HC is blocked, and the HC lifetimes therefore tend to approach to similar values for NCs with different defect densities, as shown in Fig. 4b. This effect of trap filling at higher excitation fluence is further elaborated in the later section.

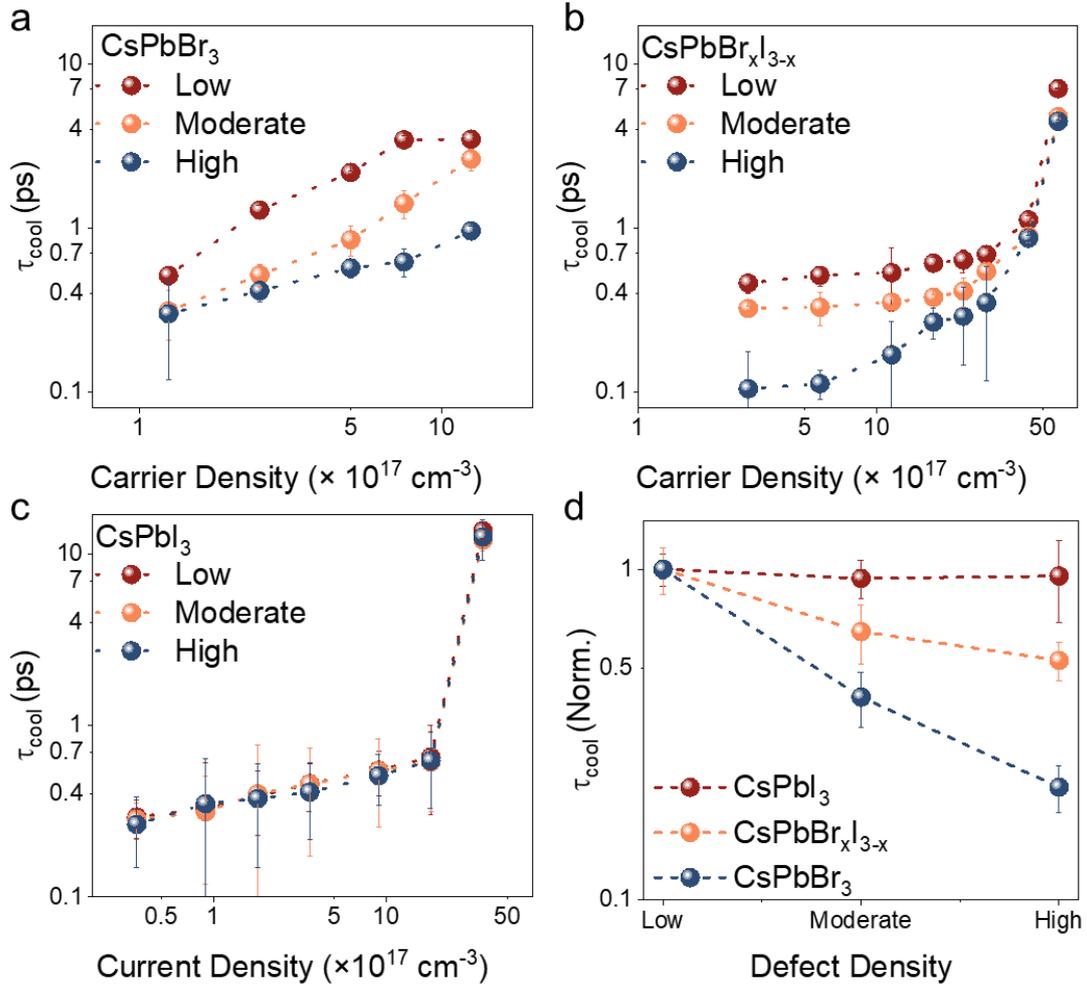

**Fig. 4 | HC lifetime ($\tau_{cool}$) fitted from pump-probe transient absorption spectroscopy measurements** of **a,** $CsPbBr_3$, **b,** $CsPbBr_xI_{3-x}$, and **c,** $CsPbI_3$ NCs with different defect densities. **d,** Hot carrier lifetime against defect densitie of perovskite NCs. The normalised $\tau_{cool}$ for $CsPbBr_3$ is based on the relaxation lifetime at 194.3 μJ cm$^{-2}$ pulse$^{-1}$ fluence (carrier density of 12.5×10$^{17}$ cm$^{-3}$), for $CsPbBr_xI_{3-x}$ at 178.2 μJ cm$^{-2}$ pulse$^{-1}$ fluence (carrier density of 43.5×10$^{17}$ cm$^{-3}$) and for $CsPbI_3$ is at 193.4 μJ cm$^{-2}$ pulse$^{-1}$ fluence (carrier density of 16.6×10$^{17}$ cm$^{-3}$).

**Influence of traps on hot carrier cooling lifetime probed by three-pulse TAS**

In the two-pulse PP approach described thus far, all charge-carriers (hot and cold) are formed by a single ('pump') excitation event. As such, the density-dependence of the intrinsic HC lifetime (including the effect of trapping) can be obscured by the 'Auger re-heating' effects at high excitation fluences. Moreover, in PP-TAS, the HCs are initially formed with different excess energies due to the variations in band gap between different NC compositions. Therefore, to obtain a more quantitative description of the effect of traps, we employed a three-pulse pump-push-probe (PPP) spectroscopic approach, in which an additional narrowband near-IR push pulse is introduced after the pump to optically re-excite the band-edge (cold) carriers to a higher energy (hot) state. The working principle of the PPP-TA setup is shown schematically in Fig. 5a, with specific details in Methods. A 1300 nm wavelength push is chosen, as this corresponds to a broad featureless intraband PIA for these systems, as shown in Supplementary Fig. 10, and described in detail elsewhere[40,41]. To ensure the push exclusively acts on cold carriers and the resulting excess energy is the same across all systems irrespective of their band gap, we introduce the push beam 10 ps after the initial 400 nm pump excitation.

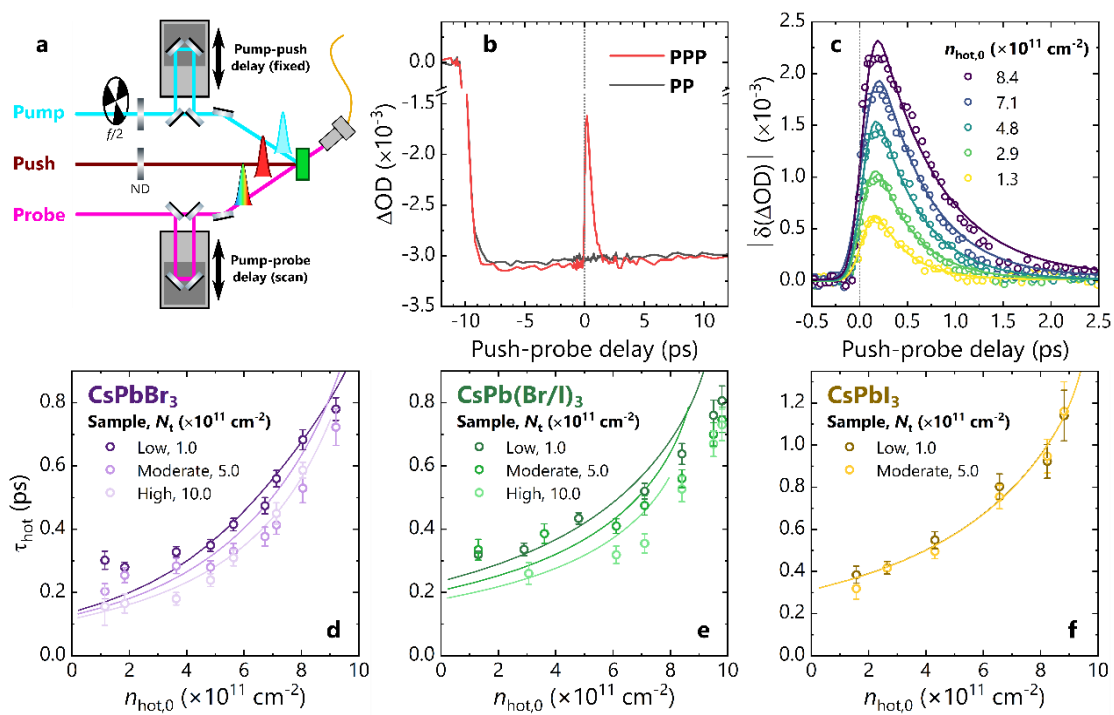

**Fig. 5 | Hot carrier cooling kinetics based on pump-push-probe transient absorption spectroscopy. a,** Schematic diagram of the pump-push-probe TA setup used. **b,** Comparison of the pump-probe and pump-push-probe ground state bleach decay kinetics with a pump-push delay of 10 ps for low defect density $CsPbBr_3$ NC solution. **c,** Hot carrier density (or push-fluence) dependent GSB decay curves. HC lifetimes obtained from fitting the pump-push-probe TA measurements for **d,** $CsPbBr_3$, **e,** $CsPbBr_xI_{3-x}$, and **f,** $CsPbI_3$ NCs, respectively. Solid lines are results from the kinetic model described later.

To verify our hypothesis that the near-IR push pulse exclusively acts to create a transient hot state, we first discuss its effect on the broadband visible TA signals of $CsPbBr_3$ NCs. As can be seen from Supplementary Fig. 11, notable differences in all TA features are observed immediately following the arrival of the push: the magnitudes of the GSB and above-gap PIA are momentarily reduced, concomitant with the reappearance of the sub-gap PIA. Each of these features are also seen upon above-gap excitation in the above two-pulse experiments.

Representative GSB kinetics obtained based on CsPbBr$_3$ NCs under the same pump fluence in both PP and PPP approaches are displayed in Fig. 5b. The introduction of the push momentarily depletes the GSB as cold carriers are re-excited to hot states, before the GSB recovers to the PP baseline commensurate with HC cooling. As shown in Fig. 5c, increasing the push fluence increases the density of HCs formed, and therefore also the magnitude of this initial change ($\delta\Delta$OD) in the GSB. Fitting this push-induced bleach recovery with a Gaussian-convolved mono-exponential decay yields the HC lifetime ($\tau_{hot}$). A low pump fluence is used in all PPP measurements (such that the average carrier density per NC $\langle N \rangle \ll 1$), and the HC lifetimes extracted as a function of initial HC density ($n_{hot,0}$). That is, the hot phonon bottleneck effect can thus be isolated within a total carrier density that precludes competing Auger re-heating effects. We note that identical HC lifetimes are extracted from the kinetics of the GSB and both PIAs (Supplementary Fig. 11), but the GSB is analysed in all cases hereinafter due to its greater signal amplitude.

The HC lifetime for all three halide compositions, as well as their respective purified samples, are plotted as a function of HC density in Fig. 5d-f. The fitted HC lifetimes obtained from PPP-TA follow similar trends to the PP-TA results described earlier. In all cases, there is a general increase in $\tau_{hot}$ with increasing HC density, which is consistent with the hot phonon bottleneck effect.

Although the general trend of an increase in HC lifetime with an increase in HC density is observed across all the NC compositions investigated, systematic differences are observed between the pristine and defective samples. We rationalize this observation as follows. It is clearly evident from Figs. 5d-e that the intrinsic HC lifetimes for the defective Br- and Br/I-based NCs are shorter than their respective pristine samples. This points to halide vacancy-related defects in CsPbBr$_3$ and CsPbBr$_x$I$_{3-x}$ systems introducing a competing channel for the intraband relaxation process, leading to faster dynamics. Under the highest density of HCs, the differences in HC lifetime between pristine and defective samples are diminished. We show below that this likely arises due to trap state saturation: when all of the vacant trap sites are filled, the competing pathway for intraband relaxation processes is blocked and therefore the HC lifetime approaches that of their respective pristine samples. In the CsPbI$_3$ NCs, no differences are observed between the pristine and defective samples in spite of the introduction of halide vacancies apparent from compositional analysis (Supplementary Figs. 2a-c), confirming that defect tolerance is retained for HCs in this system.

To illustrate and quantitatively formulate the above hypothesis regarding the various HC relaxation routes, particularly carrier-trap and carrier-phonon interactions, a simple kinetic model was developed based on Eq. 3:

$$\frac{dn_{\text{hot}}}{dt} = \tilde{I}(t)n_{\text{cold}} - \alpha n_{\text{cold}}n_{\text{hot}} - \phi n_{\text{ph}}n_{\text{hot}} - \gamma(N_t - n_t^*)n_{\text{hot}} \qquad (3)$$

where $\tilde{I}$ represents the push Gaussian envelope. $n_{\text{hot}}$, $n_{\text{cold}}$, $n_{\text{ph}}$, $N_t$ and $n_t^*$ are the densities of hot and cold carriers, vacant phonons, and total and occupied traps, respectively. The coefficients $\alpha$ and $\phi$ represent hot-cold carrier and carrier-phonon scattering, and are based on values extracted from our previous work[42-44], while $\gamma$ denotes the propensity for these defects to act as traps for HCs. The full system of coupled differential equations can be found in the Supporting Information Eq. S6-9. The modelled results can be found overlaid on the experimental data in Fig. 5d-f, and show good agreement with physically-reasonable fitting parameters (Supplementary Table 5). In the modelling, $N_t$ is mainly governed by the number of purification steps. As $N_t$ increases, the cooling lifetime is shortened in the Br-containing NCs, for which $\gamma$ values are non-zero. The effect of defect densities on HC cooling can also be influenced by the push fluence due to the trap filling effect, *i.e.*, as $n_t^* \to N_t$, the trapping term in Eq. 3 approaches zero. This explains the observed up-curvature for the purified CsPbBr$_3$ and CsPbBr$_x$I$_{3-x}$ NCs. Meanwhile, $\gamma = 0$ in CsPbI$_3$, leading to negligible differences between purified and pristine NCs. As such, the numerical model compares very satisfactorily with salient trends observed in experimental Figs. 5d-f, giving further credence to the notion that trapping can accelerate HC loss. Concretely, HCs undergo direct trapping – rather than sequential cooling then trapping, which would not manifest as a change in HC lifetime – and the driving force for this process is the energy offset to deep traps in the defect-intolerant NCs. Furthermore, these results highlight more generally that an interplay of the hot phonon bottleneck and trap state saturation could lead to longer HC lifetimes in the defective samples.

**Conclusion**

In conclusion, we found that the defect tolerance of band-edge carriers in lead-halide perovskites is a good predictor for that of HCs. HCs are lost due to non-radiative recombination more readily in the wide-gap CsPbBr$_3$ and CsPbBr$_x$I$_{3-x}$ perovskites than in the narrow-gap CsPbI$_3$ perovskites. After intentionally introducing additional defects to the wide-gap NCs, faster HC relaxation is observed due to the reduction of the hot phonon bottleneck and Auger reheating effects. We posit that HCs can therefore trap directly, without needing to first cool to the band-edge. The finding suggests that HCs are not universally defect-tolerant in CsPbX$_3$ perovskites, but rather depend on trap depth, which may help to explain the current conflicting conclusions in the literature. Our work implies that there is a greater chance of realising HC solar cells or lasing from iodide-based perovskites, and that these devices can tolerate the extra defects that could be generated during the ligand exchange needed to improve the macroscopic charge-carrier transport of the NC film. We expect that these results can be more widely generalised beyond the inorganic halide perovskite system we investigated here towards hybrid systems as well. Our results suggest that narrow-gap formamidinium-based perovskites, found to exhibit greater defect tolerance than methylammonium-based perovskites[45], are promising for further investigation.

**Acknowledgements**

J.Y. and R.L.Z.H. acknowledge support from a UK Research and Innovation (UKRI) Frontier Grant (no. EP/X029900/1), awarded via the European Research Council Starting Grant 2021 scheme. J.Y. also gives thanks to Cambridge Philosophical Society for the Research Studentship Grant and Churchill College for various travel and research grants. N.M. and A.A.B. acknowledge support from the European Commission through the Marie Skłodowska-Curie Actions Project (PeroVIB, H2020-MSCA-IF-2020-101018002). Y. Z. acknowledges funding from the National Natural Science Foundation of China under Grant No. 12304036, the Open Project of Guangdong Provincial Key Laboratory of Magnetoelectric Physics and Devices (No. 2022B1212010008), the Guangdong Basic and Applied Basic Research Foundation (2023A1515010071), the Guangzhou Basic and Applied Basic Research Foundation (SL2022A04J00048), and the Fundamental Research Funds for the Central Universities, Sun Yat-sen University (23xkjc016L.D. thanks the Cambridge Trusts and the China Scholarship Council for funding. ). L.v.T. thanks the Winton Programme for the Physics of Sustainability and the Engineering and Physical Sciences Research Council for funding. A.A.B. acknowledges support from the Royal Society and Leverhulme Trust. R.L.Z.H. thanks the Royal Academy of Engineering through the Research Fellowships scheme (no. RF\201718\17101), as well as the Centre of Advanced Materials for Integrated Energy Systems (CAM-IES; EPSRC grant no. EP/T012218/1). The authors thank Dr. Mark Isaacs for XPS measurements. The author also would like to acknowledge that the X-ray photoelectron (XPS) data was acquired at the EPSRC National Facility for XPS ("HarwellXPS", EP/Y023587/1, EP/Y023609/1, EP/Y023536/1, EP/Y023552/1 and EP/Y023544/1).'

**Author contributions**

J.Y. and R.L.Z.H conceived the project. J.Y., L.D. and C.O-M synthesized and purified the nanocrystals with help from L.P.; J.Y. conducted pump-probe transient absorption measurement with helps from L.D. and P.G.; N.M. conducted the pump-push-probe transient absorption measurements for nanocrystals with different compositions and defect densities and analysed the data. B.P.C performed the detailed kinetic modelling under the supervision of A.B.; Z.C. did the DFT calculation under the supervision of Y.Z.; J.Y. and X.F. performed energy dependent PLQY measurement. J.M. helped with TCSPC measurements under the supervision from S.D.S.. L.v.T. and Z.Y. performed TEM measurements. A.R. and R.L.Z.H. supervised the work. All authors contributed to writing and editing the manuscript.

**Conflicts of Interest**

Authors declare no competing interests.